\documentclass[aps,prl,twocolumn]{revtex4}
\usepackage{amsmath}
\usepackage{amsfonts}
\usepackage{amssymb}
\usepackage{graphicx}
\begin{document}
\date{21/10/2014}
\title{Modeling of evolution of a complex electronic system to an ordered hidden state: application to optical quench in \emph{1T}-TaS$_{2}$.}
\author{S. Brazovskii}
\affiliation{LPTMS-CNRS, UMR8626, Univ. Paris-Sud, Bat. 100, Orsay, F-91405
France} \email{brazov@lptms.u-psud.fr}
\affiliation{Dept. of Complex Matter, Jozef Stefan Institute, Jamova 39, SI-1000, Ljubljana, Slovenia}
\affiliation{International Institute of Physics, 59078-400 Natal, Rio Grande do Norte, Brazil}
\begin{abstract}
Femto-second techniques addressing  phase transitions induced by optical pumps have allowed recently to put an ambitious goal to attend hidden states which are inaccessible and even unknown under equilibrium conditions. Recently (*),  the group from Slovenia led by D. Mihailovic  achieved a bistable switching to a hidden electronic state  in \emph{1T}-TaS$_{2}$. The state is stable until an erase procedure reverts it to the thermodynamic ground state.
A notoriously intricate nature of this material requires to consider simultaneous evolution of electrons and holes as mobile charge carriers, and crystallized electrons  modifiable by intrinsic defects (voids and interstitials); all that on the CDW background. Our model (*) considers mutual transformations among the three reservoirs of electrons, together with the heat production, which are dictated by imbalances of three partial chemical potentials. The phenomenological approach sheds a light on a very complicated and not yet resolved physics of this material which includes interplaying effects like CDW, Wigner crystal, commensurability, polarons, and Mott state.
\medskip\\
 *) L. Stojchevska, I. Vaskivskyi, T. Mertelj, P. Kusar, D. Svetin, S. Brazovskii, and D. Mihailovic, Science, \textbf{344}, 177 (2014); arXiv:1401.6786v3.
\end{abstract}
\maketitle
\section{Introduction: femto-hidden.}
 Femto-second pump\&probe technique uses a sharp laser pumping to a very high (10\% or more) concentrations of excitations (electron-hole pairs or bound excitons); then it follows the evolution, with a resolution down to tenths of fs, by means of fast probes: optics, ARPES, diffraction.
There are several goals and advantages: disentangle fast ($10^{1-2}$fs) electronic and slower (ps) lattice degrees of freedom to understand properties of the starting system \cite{peterson,perfetti,tPES}, provoke the dynamic evolution over a wide range of the phase diagram \cite{NaturePhys}, reach short leaving metastable cooperative states unavailable in thermodynamic approaches, to obtain a truly stable hidden (H) state with an unlimited life time. The last goal has been just achieved \cite{science}, for the first time in an electronic system, which theoretical description gives a content to this article. Properties of the \emph{H} state are markedly different from any other state of the system: it exhibits a large drop of electrical resistance, strongly modified single particle and collective mode spectra.

 \section{Complexity of \emph{1T}-TaS$_{2}$.}
 \emph{1T}-TaS$_{2}$ is a rich system which exhibits multiple forms of the charge ordering already under equilibrium conditions. Further nearby equilibrium states are revealed upon application of the external pressure \cite{TaS2-SC} or by doping \cite{TaS2+Fe-SC}, both of which make \emph{1T}-TaS$_{2}$ superconducting.
First we shall summarize  microscopic foundations (see \cite{Tosatti,Nakanishi,bandsRossnagel,bandsFreericks}) adding some our disputable view provoked by the H state problem.

\begin{figure}[h]
  \includegraphics[width=\columnwidth,height=4cm]{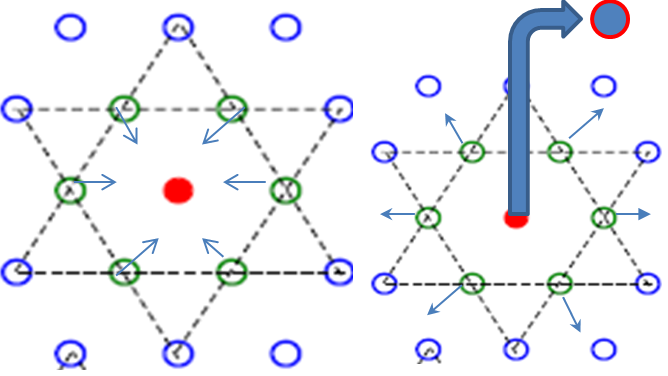}\\
  \caption{Polaronic deformations around the occupied Ta atom and their release after the electron is removed.}
\label{fig:TaS2-star+relax}
\end{figure}

The background of electronic events in \emph{1T}-TaS$_{2}$ is a narrow-gap insulator ( the gap $Eg=\Delta_e+\Delta_h$ is $\approx$ 0.6eV as from the STM \cite{kim} or 0.8eV from optics \cite{gasparov,optics-velebit}, see Fig.\ref{fig:bands+scheme}). It is formed from the parent metal (with one d-electron per Ta site) by the high-temperature ($T_c>500K$) CDW. An incomplete nesting of the CDW leaves each 13-th electron ungaped which in a typical CDW would give rise to a pocket of carriers. But here, each excess carrier is self-trapped by inwards displacements of the surrounding atomic hexagon forming the "David star" unit -  Fig.\ref{fig:TaS2-star+relax} which gives rise to the intragap local level (the red line in Fig.\ref{fig:bands+scheme} accommodating this electron.
These charged heavy polarons inevitably arrange themselves into the Wigner crystal (WC) which is always a trigonal super-lattice; here by Nature chance it is formed upon the underlying hexagonal lattice. This ideally locked form is the commensurate (C) phase. This might be the Mott insulator (for fast processes leaving the lattice intact) because adding another electron to the split-off level will require for the Hubbard repulsion energy U which is of the same order as  $E_g$. The trigonal arrangement is frustrating for spins thus preventing their AFM ordering  which makes it an ideal stage for the Mott insulator (MI) \cite{Tosatti}. Position of the upper Hubbard band (UHB) with respect to the electronic band bottom at $\Delta_e$ and their hybridization is an unresolved issue - they seem to be almost degenerate \cite{bandsRossnagel,bandsFreericks}. Another big unknown is the 3D coordination of David stars which are thought sometimes to form synchronous lines in the interplane direction. That would give rise to a strong overlap of 13'th electrons if they are formed by the Ta $d_{z^2}$ orbitals \cite{bandsFreericks} or not strong at all if their nature are $d_{x,y}$ orbitals \cite{bandsRossnagel}.

Exciting the self-trapped electron from the intragap level deprives the inwards deformations from reasons of existence, the David star smooths out in favor of a void in the WC while the electron is sent to the higher delocalized band or recombines with the band hole produced by the optical pumping, Fig.\ref{fig:TaS2-star+relax}. Observation of the mid-gap spectral feature following the sub-threshold photoexcitation \cite{perfetti,polarons,tPES} is consistent with the transient change of polaron density.
Adding an electron to the UHB may form a doubly occupied site (with the enhanced deformation that will be the  bipolaron) but the Coulomb energy cost most probably will not allow for its stability. Mitosis into two polarons will proceed thus creating an interstitial in the WC; both voids and interstitials have been seen by the STM \cite{STM}. If the defects are not expelled, then the WC melting gives rise to the incommensurate (IC) phase seen at higher T. At lower T, the phase separation takes place: voids are expelled in-between dense C clusters  yielding the near-commensurate (NC) phase. Such a self-tuned Wigner-Mott state stays always, even under doping,  in half-filled MI regime: number of sites is regulated to adjust to the number of electrons which is a common notion in CDWs \cite{braz:2007}. There is no such a thing here as the doped MI.

\section{Equilibria.}
\begin{figure}[h]
  \includegraphics[width=\columnwidth,height=4cm]{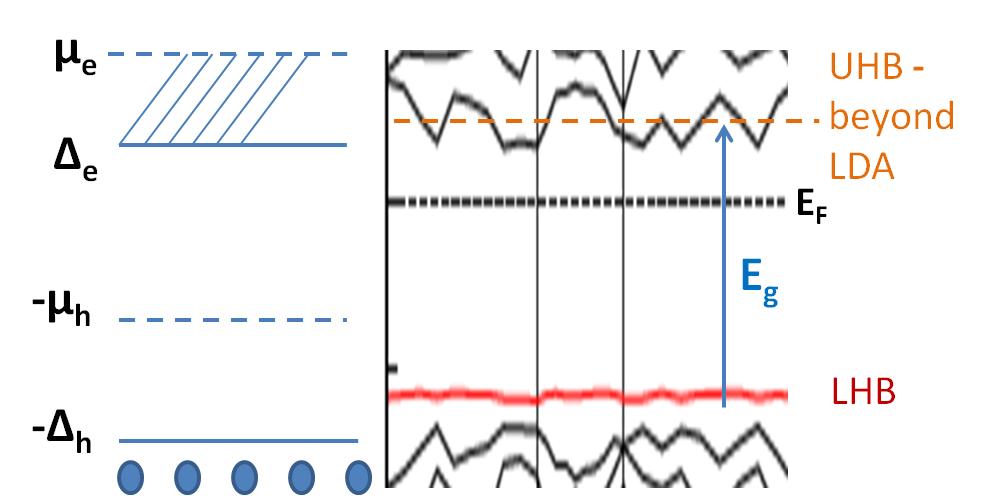}\\
  \caption{Example: $\mu_e>\Delta_e$ – degenerate Fermi gas of electrons
$|\mu_h|< \Delta_h$ - thermal activation, Boltzmann gas of holes. This drawing includes results from \cite{bandsFreericks}.}
\label{fig:bands+scheme}
\end{figure}

\begin{figure}[h]
  \includegraphics[width=\columnwidth,height=4cm]{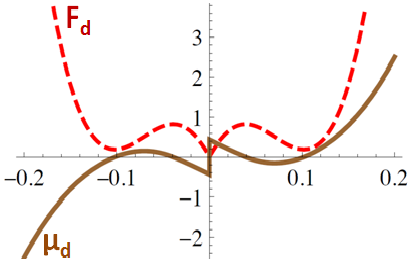}\\
  \caption{The CDW free energy $F_d$ (dashed line) and the chemical potential $\mu_d=dF_d /dn_d$  as functions of defects concentration $n_d$ =\#voids minus \#interstitials.}
\label{fig:Fd+mud}
\end{figure}

Equilibrium states and rates of equilibration are determined by the game of partial chemical potentials for each reservoir of electrons. Band states near rims of the band gap are considered as free 2D electrons, then
\begin{equation}
\mu_{e,h}(n)=\Delta_{e,h}+T\ln(\exp(n_{e,h}/(TN_{e,h})-1).
\label{mu-eh}
\end{equation}
where $N_{e,h}$ are densities of states, Fig.\ref{fig:bands+scheme}.
For defects, we adopt the scheme of incommensurable 2D superlattices \cite{McM}. To give rise to the experimental 1st order C-NC transition, the free energy $F_d(n_d)$ as a function of defects' concentration $n_d$ could be taken as any double-well curve  or as its numeric precision from semi-phenomenological studies \cite{Nakanishi}. Our choice is a qualitatively transparent theory \cite{bak} where the 1st order character of the transition is explained as effect of crossings of domain walls. (Still, recall an incompleteness of  typical old theories of commensurability in in (quasi) 2D systems in applications to electronic superstructures: no Coulomb energy from charged walls, no constraints on conservation of electrons.)
Assuming the symmetry between vacancies and interstitials, we choose the parametrization

\begin{equation}
F_{d}(n_{d})=E_{DW}(C_{0}|n_{d}|+C_{1}|n_{d}|e^{-1/(\xi|n_{d}|)}
-C_{2}\xi n_{d}^{2}+C_{4}\xi^{3}n_{d}^{4})\label{eq:F-d}
\end{equation}
where $C_{n}$ are numeric constants, $\xi$ is the domain wall width, $E_{DW}$ is its energy scale per a constituent defect. The first two terms are standard for a picture of the 2nd order $C$-$IC$ transition: in thermodynamic equilibrium, the coefficient $C_{0}<1$, as a function of $T$, reduces the DW energy, then for $C_{0}<0$ the walls start to be created but their concentration is stabilized by the repulsion coming from the term $\sim C_{1}$.
The term $\sim -C_{2}$ appears for non-collinear arrays of domain walls
which now intersect in points with a concentration $\sim n_{d}^{2}$.
This energy is expected to be negative, implying $\sim C_{2}>0$ in (\ref{eq:F-d}). Together with the last stabilizing term $\sim C_{4}$ to take into account the repulsion between the crossings, we obtain the desired non-monotonous curve for $F_{d}$ and its derivative $\mu_d=dF_d /dn_d$ - the chemical potential of defects as shown in Fig. \ref{fig:Fd+mud}.

\begin{figure}[h]
  \includegraphics[width=\columnwidth,height=4cm]{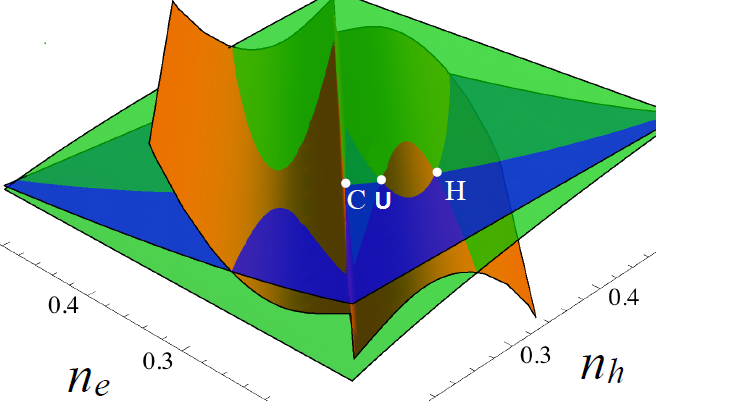}\\
  \caption{Chemical potential surfaces $-\mu_e(n_e)$ (blue), $\mu_h(nh)$ (green), $\mu_d(n_e-n_h)$ (orange).
C,U,H are triple intersections where $-\mu_e(n_e)= \mu_h(n_h)=\mu_d(n_e-n_h)$.
C and H are points of local equilibria under the constraint $n_e-n_h = n_d$.
In the C point,  $n_e=n_h , n_d=0$; point U of the energy maximum is repulsive.}
\label{fig:potentials}
\end{figure}

The three surfaces of chemical potentials $-\mu_{e}$, $\mu_{h}$ and $\mu_{d}$ as functions of $n_{e}$, $n_{h}$ and $n_d=n_e-n_h$ respectively are shown in Fig. \ref{fig:potentials}.
($\mu_h$ and $\mu_d$, as both counting the electron's deficiencies, should be always taken with an opposite sign with respect to $\mu_e$.)

\begin{figure}[h]
  \includegraphics[width=0.49\columnwidth,height=4cm]{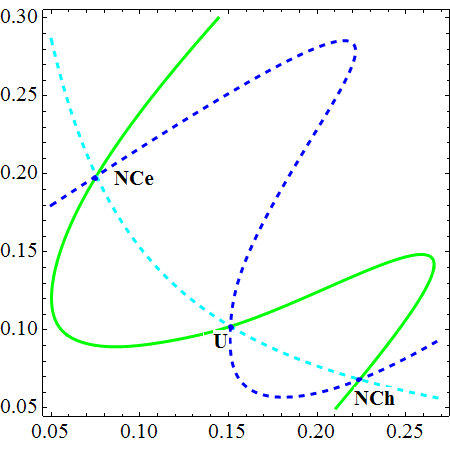}
  \includegraphics[width=0.49\columnwidth,height=4cm]{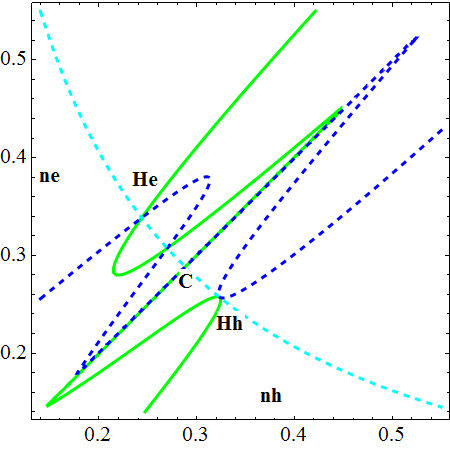}
  \\
  \caption{Lines of partial equilibrations - crossings of surfaces shown in Fig.\ref{fig:potentials}: $-\mu_e(n_e)=\mu_h(nh)$ (cyan dashed), $\mu_d(n_e-n_h)=\mu_h(nh)$ (green), $\mu_d(n_e-n_h)=-\mu_e(n_e)$ (blue dashed).
  Left panel: for these parameters there are two NC phases, enriched by either electrons ($NC_e$) or holes ($NC_h$, which is just appearing), and there is no stable C phase.
  Right panel: the C phase and two allowed H phases enriched by electrons ($H_e$) or holes ($H_h$). For yet different parameters (not shown) there may by no crosssections at all among blue and green lines, then only the C phase would exist.}
\label{fig:lines}
\end{figure}

Variations of parameters and temperature can change the plot's topology as we see in Fig.\ref{fig:lines}.

\section{Kinetics.}
To model the nonequilibrium evolution, we need to consider the relaxation kinetics between the three reservoirs. Mutual transformations among the reservoirs, together with the concomitant heat production, are dictated by imbalances of three partial chemical potentials $\mu_j$. The rates equations are chosen in a simplest form satisfying to the condition that the exchange rate among any two reservoirs vanishes when the corresponding chemical potentials become equal $\delta\mu_{i,j}=0$.
Finally, kinetic equations for the time evolution of $n_{h}(t)$ and $n_{e}(t)$ acquire the form:

\begin{eqnarray}
\frac{dn_{h}}{dt}=-k_{eh}n_{e}n_{h}(\mu_{e}+\mu_{h})-k_{hd}n_{h}(\mu_{h}-\mu_{d}) & +P(t)
\label{eq:rates1}
\\
\frac{dn_{e}}{dt}=-k_{eh}n_{e}n_{h}(\mu_{e}+\mu_{h})-k_{ed}n_{e}(\mu_{e}+\mu_{d}) & +P(t)
\label{eq:rates2}
\end{eqnarray}
where $k_{ij}$ are the coefficients of the recombination rates after extracting dependencies on $n_{i}$ and $\delta\mu_{i,j}$, $P(t)$ is the temporal profile of particles production by pumping.

The temperature evolution is taken into account by the energy balance equation. An advantage of our accent upon chemical potentials is that we know the power released by processes of electronic transformations:
\begin{equation}
J=\sum_{j}\mu_{j}(dn_{j}/dt),
\label{eq:temp}
\end{equation}
 Furthermore, T  branches to the electronic $T_e$ for $\mu_e$ , $\mu_h$ and the lattice one $T_l$ for $\mu_d$.

\begin{figure}[h]
\includegraphics[width=\columnwidth,height=4cm]{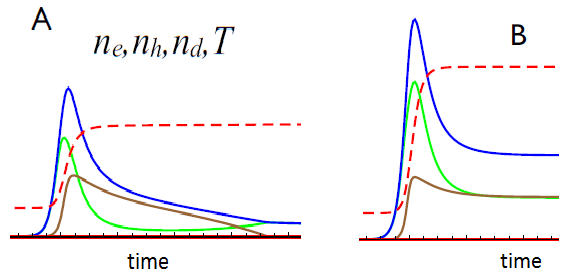}
\caption {The calculated $n_{e}$ (blue), $n_{h}$ (green), $n_{d}$ (brown),
and $T$ (red dashed) as functions of time:
A) below threshold ($U<U_T$) and B) above threshold ($U>U_T$).}
\label{fig:plots}
\end{figure}

\begin{figure}[h]
\begin{centering}
\includegraphics[width=\columnwidth]{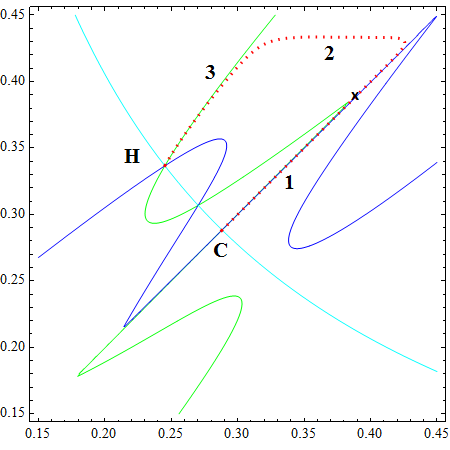}
\end{centering}
\caption{Trajectory - parametric plot of $n_e(t)$ and $n_h(t)$ at T=cnst (red dots) superimposed on lines of partial equilibriums given by intersections $\mu_e =- \mu_d$ - blue, $\mu_h = \mu_d$  - green, $\mu_e = -\mu_h$ - cyan. The three segments (1-3) of the trajectory are explained in the text.
}
\label{fig:traj123}
\end{figure}

The formulas (\ref{eq:rates1},\ref{eq:rates2},\ref{eq:temp}) form a complete set of equations governing the system evolution. Their solution yields partial time dependencies in Fig. \ref{fig:plots} and finally the trajectories of the Fig. \ref{fig:traj123}, .

The calculation does not extend to the cooling down to the temperature of the  environment, so T remains high after all the pulse energy is transferred.  $T_e$ shows a maximum, then it decreases to meet the growing $T_l$. We see that $\mu_h <\Delta_h$ - the holes always form a Boltzman gas. The electrons form a degenerate Fermi gas only temporarily when $\mu_e >\Delta_e$; they change the activated regime at longer time; at much longer T they will become degenerate again after final cooling to the ambient T. These dependencies generate trajectories which were shown in \cite{science}, complemented also by the one for the erasing pulse converting the H phase back to the C one. Here in Fig.\ref{fig:traj123} we show a trajectory for simplified calculations without taking into account the temperature evolution, unlike plots of Fig.\ref{fig:plots}; T was kept at an intermediate T=cnst. That was done for the sake of visualization since otherwise the reference lines would change with T, hence with t, and the trajectory could not be easily interpreted.

With the system in the ground C state, the pulse was given creating an equal numbers of e and h which leads the trajectory away from the C point while still up the C line (segment 1). For a while, this bisectrix keeps a degenerate equilibrium of both e and h with respect to d. But above the level marked by "x" the green line deviates - the h-d equilibration cannot be kept any more while the equilibrium e-d still persist. At a critical concentration, the holes start to be converted into defects (voids) leaving electrons almost intact, hence the C line is left by the nearly horizontal one ($n_e\approx$cnst, valid only for the model with T=cnst) of the segment 2. The trajectory reaches another branch  of the h-d equilibrium (green line) but cannot stay their because now the electrons are not equilibrated with any other component. By processes of e-h and e-d annihilation, the trajectory descends following the h-d line (segment 3) to the H point of the stable H phase.

The moment of truth came recently when ARPES, tr-ARPES and diffraction have been applied (at SLAC, Stanford \cite{Kirchmann}) to the H phase of \emph{1T}-TaS$_{2}$. The experiments clearly show the abundance  of electrons appearing above the threshold (in the H phase) and the conclusion was given: "trARPES suggest that e-h imbalance drives switching".

The reason for the remarkable stability of the \emph{H} state over an arbitrarily long time is that it is topologically protected: when defects are ordered into the regular structure of domain walls,
the  density $n_d$ cannot change continuously, but can do so only in discrete steps, when the number of periods of the electronic crystal changes, which is quantized. Such an effect has been demonstrated in 1D CDWs  \cite{borodin,zybtsev}. This constraint can be resolved only by energetically costly proliferation of defects - the dislocation lines.
The creation and motion of such extended objects will be substantially slowed down by the presence of intersecting discommensurations and finally arrested by pinning to lattice defects.
The erase cycle may be explained by the creep of extended defects which is known to be promoted by heating above an irreversibility line characteristic of pinning phenomena.

\section{Conclusions}
Understanding the rout to the hidden phase in a still mysterious \emph{1T}-TaS$_{2}$ gives an insight to this material as a self-tuned, phase-separated, always filled, Mott state of a Wigner crystal of polarons.
Electrons and holes as mobile charge carriers experience evolution exchanging with crystallized electrons modifiable by intrinsic defects.
Mutual transformations among the three reservoirs of electrons, together with the heat production, are dictated by imbalances of three partial chemical potentials. The main nontrivial observations, namely the appearance of a switching threshold for the pump pulse fluence, its critical pulse-length dependence, the threshold temperature for the erase cycle,  and the high conductance can thus be reproduced.
Next passage will be to generalize the rate eqs. to kinetic ones to go beyond the quasi-static limitation.

\begin{acknowledgments}
This work came from collaboration at the Jozef Stefan Institute in Ljubljana, Slovenia with D. Mihailovic, T. Mertelj,  and their colleagues. There, the work was partly funded from the ERC advanced grant TRAJECTORY.
 The author acknowledges also discussions with V.V. Kabanov and N. Kirova and thanks P.S. Kirchmann for acquaintance with his unpublished results.
\end{acknowledgments}

\end{document}